\documentclass{svjour2}                    % onecolumn
\smartqed  % flush right qed marks, e.g. at end of proof
\usepackage{graphicx}
\usepackage{bm}

\usepackage{amssymb}
\usepackage{amsmath}

\newcommand{\be}{\begin{equation}}
\newcommand{\ee}{\end{equation}}

\begin{document}

\title{Mesoscopic simulation study of wall roughness effects in micro-channel flows of dense emulsions}



\author{Andrea Scagliarini     \and
        Mauro Sbragaglia       \and
        Massimo Bernaschi
}


\institute{Andrea Scagliarini \at
           Department of Physics and INFN, University of ``Tor Vergata'', Via della Ricerca Scientifica 1, 00133 Rome, Italy \\
              \email{ascagliarini@gmail.com}           %  \\
           \and
           Mauro Sbragaglia \at
           Department of Physics and INFN, University of ``Tor Vergata'', Via della Ricerca Scientifica 1, 00133 Rome, Italy \\
              \email{sbragaglia@roma2.infn.it}           %  \\
           \and
           Massimo Bernaschi \at
           Istituto per le Applicazioni del Calcolo CNR, Via dei Taurini 19, 00185 Rome, Italy \\
              \email{massimo.bernaschi@cnr.it}
}

\date{Received: date / Accepted: date}

\maketitle

\begin{abstract}
We study the Poiseuille flow of a soft-glassy material above the jamming point, where the material flows like a complex fluid with Herschel-Bulkley rheology. Microscopic plastic rearrangements and the emergence of their spatial correlations induce cooperativity flow behavior whose effect is pronounced in presence of confinement. With the help of lattice Boltzmann numerical simulations of confined dense emulsions, we explore the role of geometrical roughness in providing activation of plastic events close to the boundaries. We probe also the spatial configuration of the fluidity field, a continuum quantity which can be related to the rate of plastic events, thereby allowing us to establish a link between the mesoscopic plastic dynamics of the jammed material and the macroscopic flow behaviour.
\end{abstract}
\section{Introduction}\label{sec:intro}

Soft-glassy materials encompass a wide variety of systems such as emulsions, foams and granular media \cite{Mason96,Katgert13,Kamrin12,Larson}. The dynamics of these complex fluids is usually characterized by relatively large packing fractions, and the microscopic constituents (i.e., droplets for emulsions, bubbles for foams, etc) are {\it jammed} together so as to exhibit a yield stress $\sigma_Y$, below which the material deforms elastically and above which it flows like a non-Newtonian fluid according to a Herschel-Bulkley rheology. A challenging question concerns the role of microscopic plastic rearrangements and the emergence of their spatial correlations inducing cooperativity flow behavior at the macroscopic level \cite{Goyon08,Goyon10,Jop12,Katgert10,Bocquet09}. Such rearrangements indeed provide {\it local} fluidization zones in the system which promote agitation of their neighbours triggering more rearrangements. This cooperative behavior can be accounted for in a modification of the continuum theory describing the local flow-curve, i.e. the relation between stresses and shear-rates. Goyon {\it et al.} \cite{Goyon08} were the first to demonstrate that such a modification can explain successfully velocity profiles of concentrated emulsions observed in experiments. In particular, they introduced the concept of {\it spatial cooperativity} length $\xi$, by postulating that the fluidity$f = \dot{\gamma}/\sigma$, defined as the ratio between shear rate $\dot{\gamma}$ and shear stress $\sigma$, follows a non-local diffusion-relaxation equation when it deviates from its bulk value
\be\label{eq:fluidity}
\xi^2 \Delta f({\bm x})+ f_b(\sigma({\bm x}))-f({\bm x}) =0.
\ee
The quantity $f_b$ is the bulk fluidity, i.e. the value of the fluidity in the absence of spatial heterogeneities. The bulk fluidity depends upon the local stress only, whereas $f$ depends upon the position in space. Its value is equal to $f_b$ without the effect of cooperativity ($\xi=0$). The spatial cooperativity $\xi$ was shown to be in the order of few times the size of the elementary microstructural constituent \cite{Goyon08,Goyon10,Geraud13}. The non-local equation \eqref{eq:fluidity} has been justified \cite{Bocquet09} starting on a kinetic model for the elastoplastic dynamics of a jammed material, which takes the form of a nonlocal kinetic equation for the stress distribution function. Such model predicts nonlocal equations of the form (\ref{eq:fluidity}), plus an equation predicting a proportionality between the fluidity and the rate of plastic events $\Gamma$,
\be\label{eq:proportionality}
f = \frac{\dot{\gamma}}{\sigma} \propto \frac{\Gamma}{G_0}
\ee
where $G_0$ is the bulk elastic modulus. The above picture was later applied to other complex fluids, such as Carbopol gels \cite{Geraud13}, granular media \cite{Kamrin12 ,Amon12}, and foams in a 2d cylindrical Couette geometry \cite{Katgert10}. The fluidity model provides a convenient framework to rationalize the flow of confined complex fluids. However, at least two points remain rarely investigated. {\it First}, the relation between $f$ and $\Gamma$: although the existence of a master relationship is observed \cite{Mansard13,OURJFM}, sometimes the scaling properties found in experiments \cite{OURJFM} and theory \cite{Bocquet09} differ from those obtained by simulations \cite{Mansard13}. {\it Second}, the issue of the boundary conditions at the solid walls for the fluidity field: as a matter of fact, one has to set it as a free fit parameter in the theory, which certainly improves the agreement between the measurements and the predictions from the fluidity model, but does not provide any insight on the role of the confining walls. Only recently, Mansard {\it et al.} \cite{Mansard14} explored the role of surface boundary conditions for the flow of a dense emulsion. Both slippage and wall fluidization are shown to be affected by the roughness, which actually contributes to the activation of plastic events, especially when the characteristic size of the roughness is larger than the characteristic droplet size.  Nevertheless, only few experimental corrugations were considered, thus motivating complementary studies to assess the robustness of those findings.\\
In recent years we developed a mesoscopic approach to the soft-glassy rheology, based on a Lattice version of Boltzmann (LB) kinetic equation with competing short-range attraction and mid-range repulsion \cite{CHEM09,EPL10,SOFT12,EPL13}. The lattice kinetic model has been shown to reproduce, with a pretty good quantitative agreement, many distinctive features of soft-glassy materials, such as ageing \cite{CHEM09} and non-linear rheology \cite{EPL10}. Specifically, it allows, with an affordable computational cost, to simulate  a collection of closely packed droplets with variable polydispersity and packing fraction, under different load conditions \cite{EPL13}. Based on such an approach, hereafter we study the Poiseuille flow of concentrated emulsions above the jamming point. The material is confined in a 2d channel where one wall is decorated with a periodic array of posts with variable wavelength, width and height. Numerical simulations allow to study local fluctuations of the fluidity field and to relate them to the rate of plastic events. We wish to stress that such analysis can strongly benefit from numerical simulations because it is extremely difficult to access experimentally local hydrodynamics.\\
The paper is organized as follows: in Sec. \ref{sec:model} we recall the essential features of the fluid-dynamical model we consider for our study; in Sec. \ref{sec:results} we discuss the numerical results at changing the details of the geometrical roughness: we will first analyze the velocity profiles and then focus on the effects on the fluidity field and the rate of occurrence of T1 plastic rearrangements across the channel; conclusions and perspectives follow in Sec. \ref{sec:conclusions}

\section{Computational Model}\label{sec:model}

\begin{figure}
\begin{center}
\includegraphics[scale=0.3]{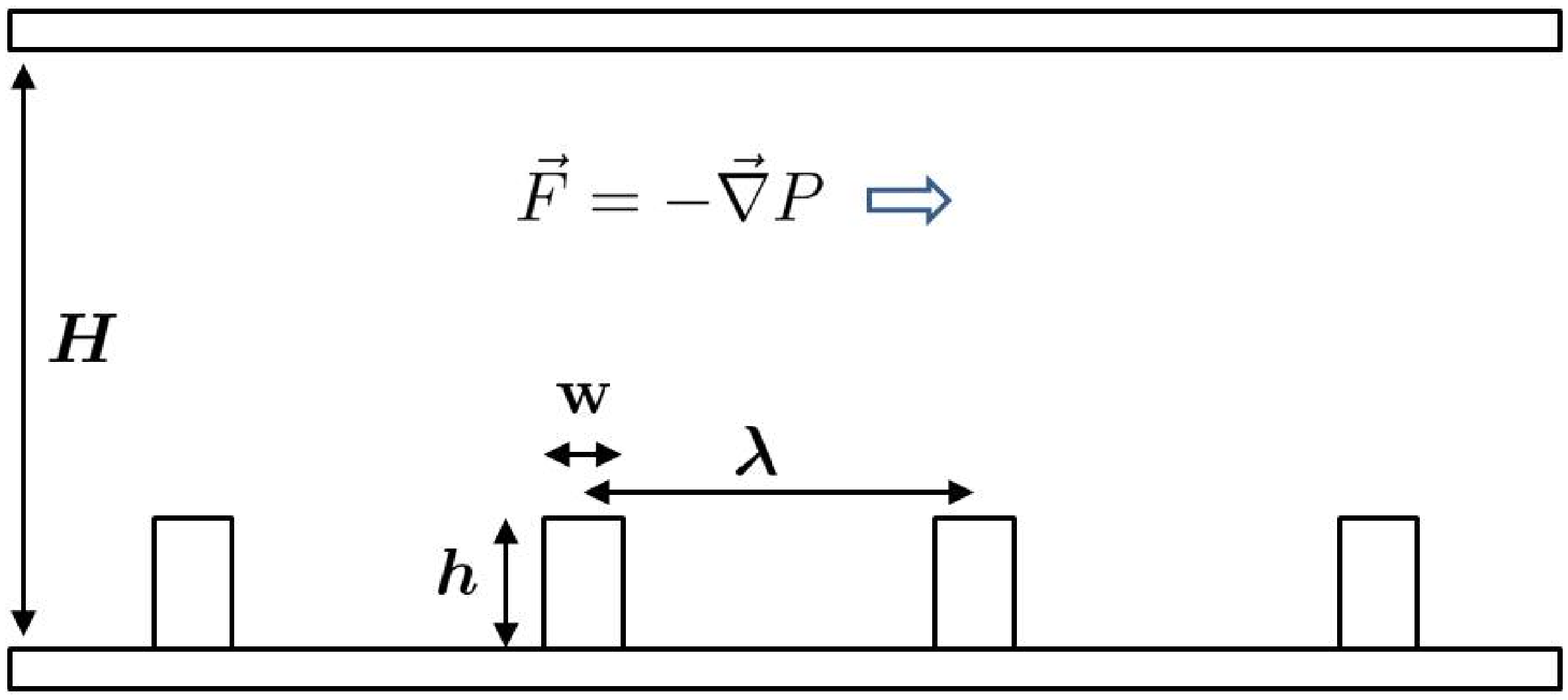}\\
\includegraphics[scale=0.47]{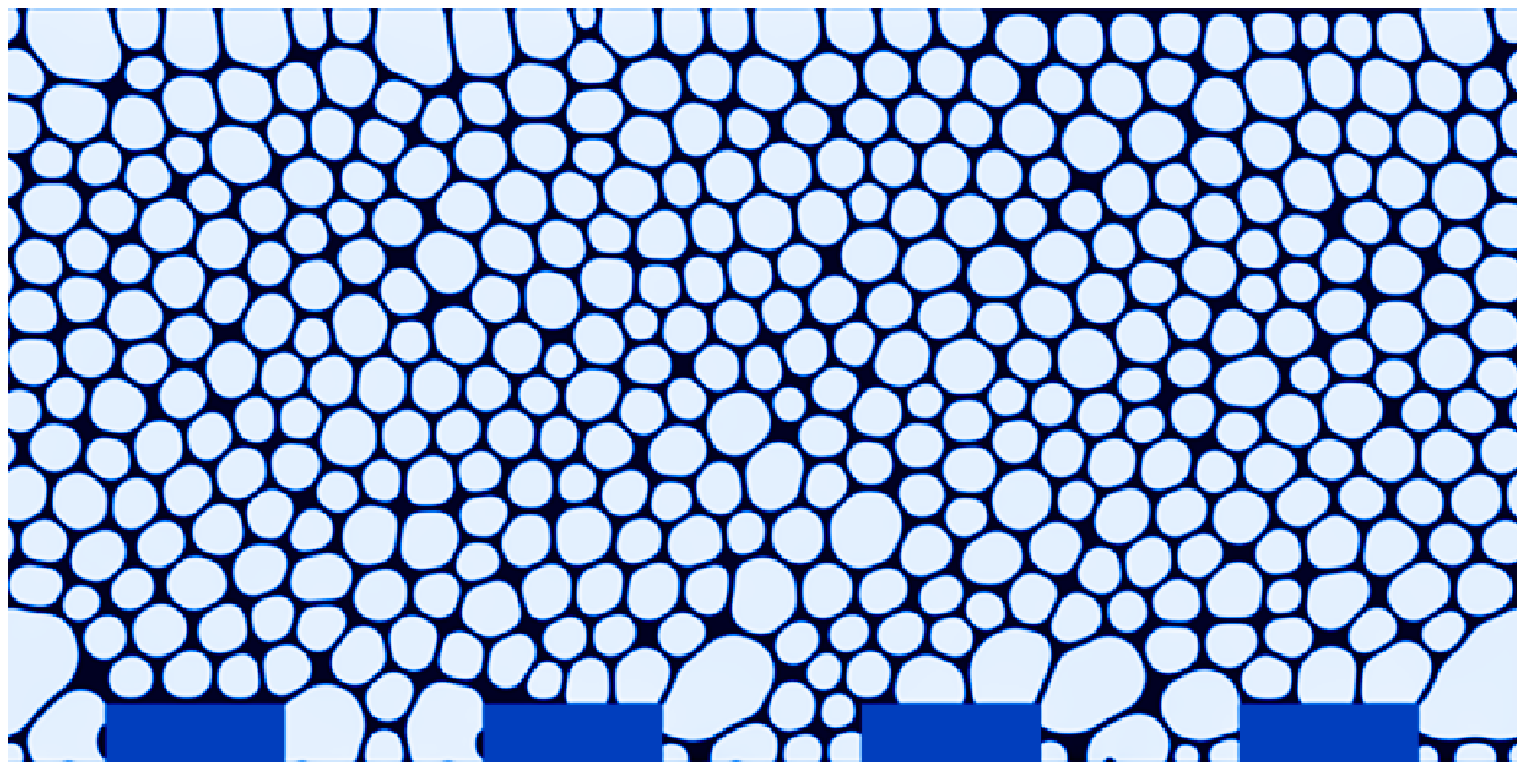}
\caption{Top Panel: view of the numerical set-up for the study of non-local effects and surface fluidization in a confined 2d soft-glassy system. Simulations are performed in a rectangular computational domain of size $L_x \times H =2 H \times H $ ($x$ is the stream-flow direction) covered by $N_x \times N_z =1024 \times 512$ lattice points. Periodic boundary conditions are applied in the stream-flow direction. The bottom wall of the channel is decorated with a periodic array of posts with wavelength $\lambda$, width $w$ and height $h$. A constant body force resulting from a  pressure gradient, $\vec{F}=-\vec{\nabla}P$, drives the system in the stream-flow direction. Bottom Panel: simulation snapshot obtained from the integration of the LB model \eqref{LB} with phase separating interactions \eqref{Phase} supplemented with competing interactions \eqref{NNandNNN}. Competing interactions \eqref{NNandNNN} are instrumental to achieve a positive disjoining pressure stabilizing thin films between neighboring droplets.\label{fig:sketch}}
\end{center}
\end{figure}

We work with a mesoscopic lattice Boltzmann model for non-ideal binary fluids, which combines a small positive surface tension, promoting highly complex interfaces, with a positive disjoining pressure, inhibiting interface coalescence. Since the model has been already described in several previous works \cite{CHEM09,EPL13}, we recall here just its basic features. The mesoscopic kinetic model considers two fluids $A$ and $B$, each described by a {\it discrete} kinetic distribution function $f_{\zeta i}({\bm x},{\bm c}_i;t)$, measuring the probability of finding a particle of fluid $\zeta =A,B$ at position ${\bm x}$ and time $t$, with discrete velocity ${\bm c}_i$, where the index $i$ runs over the nearest and next-to-nearest neighbors of ${\bm x}$ in a regular two-dimensional  lattice \cite{CHEM09}. In other words, the mesoscale particle represents all molecules contained in a unit cell of the lattice. The distribution functions evolve in time under the effect of free-streaming and local two-body collisions, described, for both fluids ($\zeta=A,B$), by a relaxation towards a local equilibrium ($f_{\zeta i}^{(eq)}$) with a characteristic time-scale $\tau_{LB}$:
\begin{equation}\label{LB}
f_{\zeta i}({\bm x}+{\bm c}_i,{\bm c}_i;t+1) -f_{\zeta i}({\bm x},{\bm c}_i;t)  = -\frac{1}{\tau_{LB}} \left(f_{\zeta i}-f_{\zeta i}^{(eq)} \right)({\bm x},{\bm c}_i;t)+F_{\zeta i}({\bm x},{\bm c}_i;t).
\end{equation}
The equilibrium distribution is given by
\be
f_{\zeta  i}^{(eq)}=w(|{\bm c}_i|^2) \rho_{\zeta} \left[1+\frac{{\bm v} \cdot {\bm c}_i}{c_s^2}+\frac{{\bm v}{\bm v}:({\bm c}_i{\bm c}_i-c_s^2 {\bf 1})}{2 c_s^4} \right]
\ee
with $w(|{\bm c}_i|^2)$ a set of weights known a priori through the choice of the quadrature \cite{Sbragaglia07a,Sbragaglia07b}. The model provides both coarse grained hydrodynamical densities, defined as $\rho_{\zeta }=\sum_i f_{\zeta i}$ and a global momentum for the whole binary mixture ${\bm j}=\rho {\bm v}=\sum_{\zeta , i} f_{\zeta i} {\bm c}_i$, with $\rho=\sum_{\zeta} \rho_{\zeta}$. The term $F_{\zeta i}({\bm x},{\bm c}_i;t)$ is just the $i$-th projection of  the total internal force which includes a variety of interparticle forces. First, a repulsive ($r$) force with strength parameter ${\cal G}_{AB}$ between the two fluids
\begin{equation}\label{Phase}
{\bm F}^{(r)}_\zeta ({\bm x})=-{\cal G}_{AB} \rho_{\zeta }({\bm x}) \sum_{i, \zeta ' \neq \zeta } w(|{\bm c}_i|^2) \rho_{\zeta '}({\bm x}+{\bm c}_i){\bm c}_i
\end{equation}
is responsible for phase separation \cite{CHEM09}. Interactions \eqref{Phase} are nothing but a lattice transcription of mean-field models for phase separation (see \cite{Bastea} and references therein). They can produce stable non-ideal interfaces with a positive surface tension \cite{CHEM09,SC}. However, they give rise {\it only} to negative disjoining pressures, i.e. ``thin'' films between neighboring droplets cannot be stabilized against rupture.  To provide an energy barrier against the rupture of such thin films, we introduce competing interactions encoding a mechanism for {\it frustration} ($F$) for phase separation \cite{Seul}. In particular, we model short range (nearest neighbor, NN) self-attraction, controlled by strength parameters ${\cal G}_{AA,1} <0$, ${\cal G}_{BB,1} <0$, and ``long-range'' (next to nearest neighbor, NNN) self-repulsion, governed by strength parameters ${\cal G}_{AA,2} >0$, ${\cal G}_{BB,2} >0$
\begin{equation}\label{NNandNNN}
\begin{split}
{\bm F}^{(F)}_\zeta ({\bm x})=&-{\cal G}_{\zeta \zeta ,1} \psi_{\zeta }({\bm x}) \sum_{i \in NN} w(|{\bm c}_i|^2) \psi_{\zeta }({\bm x}+{\bm c}_i){\bm c}_i  \\
&-{\cal G}_{\zeta \zeta ,2} \psi_{\zeta }({\bm x}) \sum_{i \in NNN} w(|{\bm c}_i|^2) \psi_{\zeta }({\bm x}+{\bm c}_i){\bm c}_i
\end{split}
\end{equation}
with $\psi_{\zeta }({\bm x})=\psi_{\zeta }[\rho({\bm x})]$ a suitable pseudo-potential function \cite{SC,SbragagliaShan11}. The pseudo-potential is taken in the form originally suggested by Shan \& Chen \cite{SC}
\begin{equation}
\label{PSI}
\psi_{\zeta}[\rho_{\zeta}({\bm x})]= \rho_{0} (1-e^{-\rho_{\zeta}({\bm x})/\rho_{0}}).
\end{equation}
The parameter $\rho_{0}$ marks the density value above which non-ideal effects come into play. The prefactor $\rho_{0}$ in (\ref{PSI}) is used to ensure that for small densities the pseudopotential is linear in the density $\rho_{\zeta}$. Elastic and viscous stresses do not enter in the simulation, but rather they {\it come out} of it; else stated, we do not impose any stress, but we ``measure'' it as an output: given the mechanical model for the lattice interactions described in (\ref{Phase})-(\ref{PSI}), an exact lattice theory is available \cite{Shan08,SbragagliaBelardinelli} which allows to connect the interaction forces to the total stress developed in the fluid (see appendix A in \cite{OURJFM}). We remark that the micro-mechanics of the model, Eqs. \eqref{Phase}-\eqref{NNandNNN}, is not meant to mimic ``specific'' physico-chemical details of a real system, but rather to model a ``generic'' soft-glassy system with {\it non-ideal fluid behavior} (e.g., non-ideal equation of state, phase separation), {\it interfacial phenomena} (e.g., surface tension, disjoining pressure) and {\it hydrodynamics} (e.g., velocity and stress fields \cite{OURJFM}).\\
In all the simulations here presented, we set the relaxation time $\tau_{LB}=1.0$ lbu (lattice Boltzmann units) in \eqref{LB}. Simulations are performed in a rectangular computational domain of size $L_x \times H =2 H \times H $ ($x$ is the stream-flow direction) covered by $N_x \times N_z =1024 \times 512$ lattice sites with periodic boundary conditions in the stream-flow direction. The top wall is just a flat wall, whereas the bottom wall is decorated with a periodic array of posts with period $\lambda$, width $w$ and height $h$ (see figure \ref{fig:sketch}). On both walls, we impose the mid-way bounce-back rule \cite{Gladrow}. As for the micro-mechanics of the model (see Eqs. \eqref{Phase}-\eqref{NNandNNN}), we fix the reference density to be $\rho_0=0.83$ lbu (LB units), and we choose phase separating interactions \eqref{Phase} with strength parameter ${\cal G}_{AB}=0.586$ lbu, corresponding to bulk densities $\rho_A=1.2$ ($0.12$) lbu and $\rho_B=0.12$ ($1.2$) lbu in the dispersed (continuous) phase. Phase separating interactions are supplemented with competing interactions having strength parameters ${\cal G}_{AA,1}={\cal G}_{BB,1}=-8.0$ lbu and ${\cal G}_{AA,2}={\cal G}_{BB,2}=7.1$ lbu. For the resolution used, this choice of the micro-mechanics ensures that thin films between neighboring droplets are just above the onset of stable interfaces. Finally, to drive the droplets in the stream-flow direction, we add a constant force $\vec{F}=-\vec{\nabla}P$ to the system.\\
For the simulations we used an enhanced version of our CUDA code described in \cite{PREGPU}. The algorithm for LB update is relatively simple to be ported on GPU. However, our code contains many distinct features with respect to other existing implementations. First of all we support a number of different boundary conditions that can be activated simply by changing a configuration file. Data structures describing the LB populations on GPU are organized according to the {\em structure-of-arrays} layout, thus allowing coalesced accesses to the global memory of the GPU. The number of threads and blocks is tunable to fit the resources available on the GPU. Each thread works sequentially on a group of lattice nodes assigned to it. For each lattice node, the thread copies data from the global memory into registers and, in the end of the update, back to the global memory. Some tests showed that using shared memory as a temporary buffer to speed-up memory operations was not only useless but actually harmful since the limited size of the shared memory imposed a reduction of the number of threads and, as a consequence, a slow down of the execution. The current version of the code makes use of a hybrid programming model that combines CUDA and non-blocking MPI primitives to run on multiple GPUs. The simulation domain is divided along the stream-flow direction in a number of ``slices'' equal to the number of available GPUs. The update of the boundaries among the slices and the exchange of the data within the boundaries among GPUs is carried out by a CUDA {\em stream} (and the CPU acting as a network co-processor of the GPU) whereas another CUDA {\em stream} carries out the update of the {\em bulk} of each subdomain. In this way there is an overlap between the boundaries exchange and the bulk update that determines a very high efficiency of the multi-GPU implementation. Most of the simulations reported in the present paper have been executed on Kepler ``Titan'' GPUs that feature 14 Streaming Multiprocessors with a total of 2688 cores running at 0.88 Ghz and have a memory bandwidth exceeding 200 GBytes/sec. The speedup with respect to a highly tuned (multi-core) CPU version is in excess of one order of magnitude.

%%%%%%%%%%%%%%%%%%%%%%%%%%%%%%%%%%%%%%%%%%%%%%%%%%%%%%%%%%%%%%%%%%%%%%%%%%%%%%%%%%%%%%%%%%%%%%%%%%%%%%%%%%%%%%%%%%%%%%%%%%%%%%%%%%%%%%%%%%%%%%%%%%%%%%%%%%%%%%%%

%%%%%%%%%%%%%%%%%%%%%%%%%%%%%%%%%%%%%%%%%%%%%%%%%%%%%%%%%%%%%%%%%%%%%%%%%%%%%%%%%%%%%%%%%%%%%%%%%%%%%%%%%%%%%%%%%%%%%%%%%%%%%%%%%%%%%%%%%%%%%%%%%%%%%%%%%%%%%%%%

\section{Results}\label{sec:results}

For studying the effect of geometrical roughness on the effective wall slip and fluidization properties of the soft-glassy material, we have performed various sets of simulations with fixed pressure gradient but variable post wavelength $\lambda$, post height $h$ and width $w$. In figure \ref{fig:sketch}, we provide a schematic sketch of the setup (top panel) and a snapshot from a numerical simulation reporting the color map of the dispersed phase (light (dark) colors indicate high (low) values of the density). Henceforth the numerical values of all lengths will be given in units of the mean droplet diameter $d$ (in these units the channel width is $H \approx 13\,d$). In the following we will first discuss how the velocity profiles are affected by various realizations of the roughness, and we will then focus on the effects on the fluidity field and the rate of occurrence of T1 plastic rearrangements \cite{WeaireHutzler} across the channel. To gain insight into the statistics of plastic events, we perform a Voronoi tessellation (by using the {\it voro++} libraries \cite{Voronoi}) of the centres of mass of the droplets so that we identify plastic rearrangements in the form of topological events, in which one edge of a given droplet collapses to zero length and neighbour droplets switching occurs.

\subsection{Velocity profiles}\label{subsec:velocity}

\begin{figure}
\begin{center}
\includegraphics[scale=0.6]{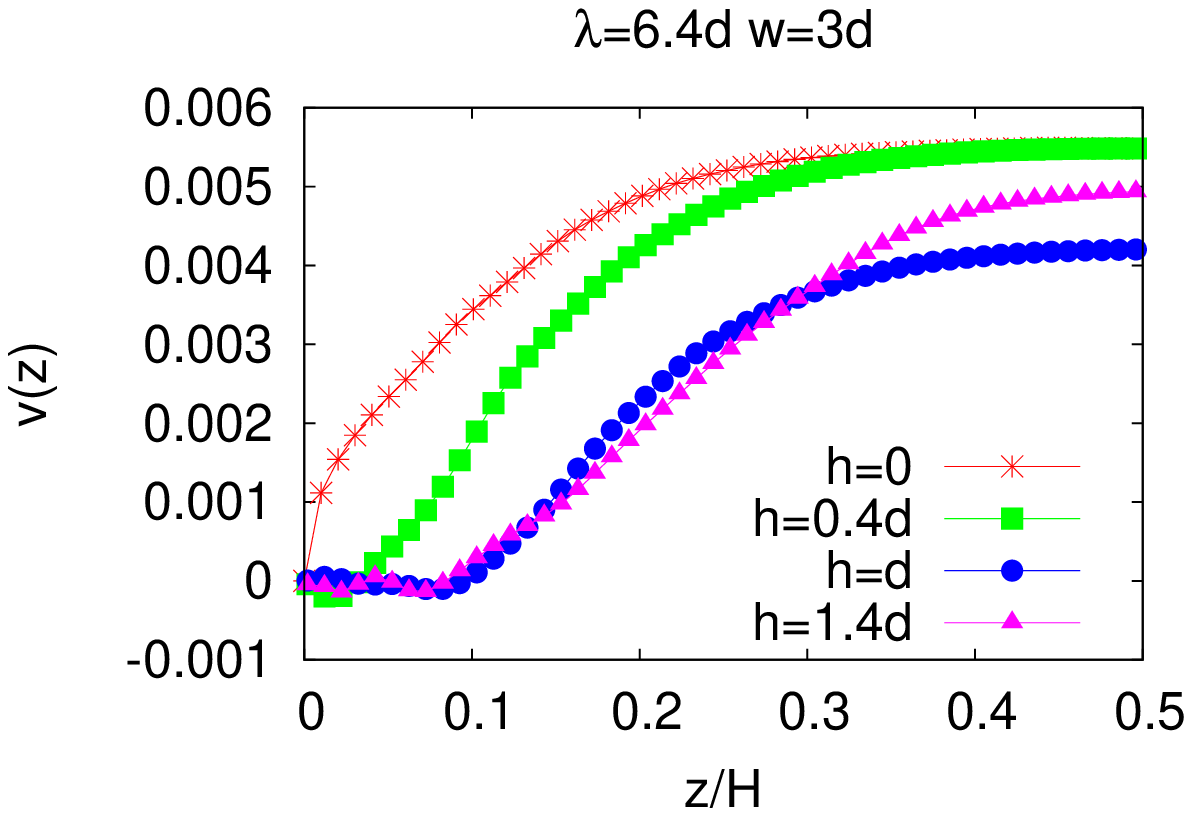}\\
\includegraphics[scale=0.6]{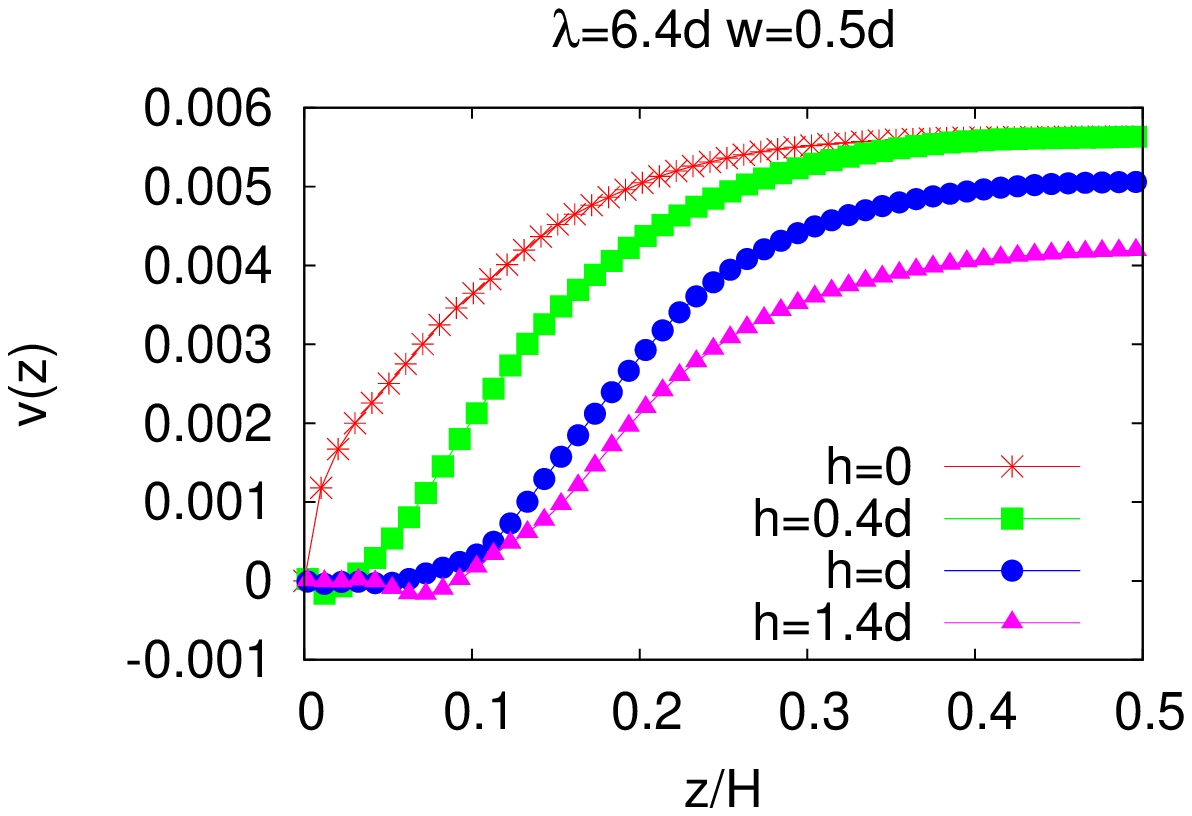}\\
\includegraphics[scale=0.6]{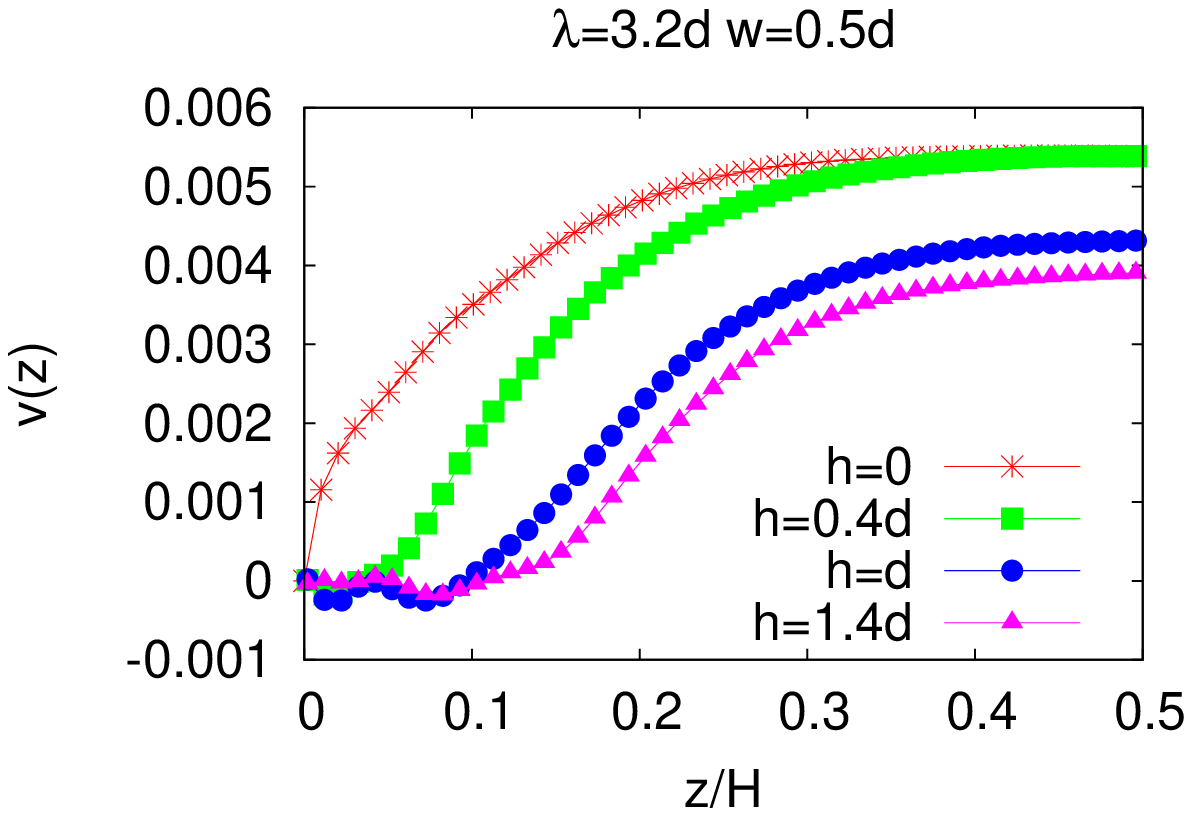}
\caption{Velocity profiles for various post heights $h=0$, $0.4\,d$, $d$, $1.4\,d$ ($d$ being the mean droplet diameter) and various couples of wavelength ($\lambda$)-width ($w$) parameters:  $(\lambda,w)=(6.4\,d, 3\,d)$ (top panel), $(\lambda,w)=(6.4\,d,0.5\,d)$ (middle panel), $(\lambda,w)=(3.2\,d, 0.5\,d)$ (bottom panel). The presence of roughness close to the bottom wall inhibits the macroscopic flow and results in lower mass flow rate through the channel. \label{fig:velocity}}
\end{center}
\end{figure}

In figure \ref{fig:velocity} we display the stream-flow time-averaged velocity profiles for the three couples $(\lambda,w)=(6.4\,d, 3\,d)$, $(\lambda,w)=(6.4\,d,0.5\,d)$, $(\lambda,w)=(3.2\,d, 0.5\,d)$, exploring three values of the post height $h=0.4\,d$, $d$, $1.4\,d$ for each of these cases. The range of posts heights is chosen so that $h$ falls both below and above the mean droplet diameter. From the figures, an interesting point of discussion emerges in attempting to determine an effective boundary condition \cite{SbragagliaProsperetti,Mansard14} close to the bottom wall. In principle, to characterize an effective boundary condition, we need to define a virtual plane at a given location $z=z_0$ inside the fluid where there is an effective slip velocity $v_{slip}$. For a given $h$, physical intuition suggests to locate such virtual plane above the bottom wall but below the posts height, i.e. $0 \le z_0 \le h$: if this is the case, we observe that the resulting effective boundary condition entails a nearly zero slip velocity, meaning that, for all the values of the roughness considered, the droplets stick to the boundaries with no net motion. This is not the case, for example, in real experiments on concentrated emulsions \cite{Mansard14}, where slip is found to emerge close to the boundaries and a suitable rescaling on the velocity profiles is needed to properly analyze data. These facts said, some common features emerge from the various velocity profiles displayed in figure \ref{fig:velocity}. Overall, it is important to observe that for $z \stackrel{>}{\sim} 4d$ the effect of roughness is lost for $h<d$ and the velocity profile of the smooth case is recovered, whereas for $h \geq d$ the velocity remains smaller than the one obtained for $h=0$. A value of $h \approx d$ can therefore be seen as a characteristic value above which the inhibition of the velocity profiles starts to matter, in line with experimental observations \cite{Mansard14}.

%%%%%%%%%%%%%%%%%%%%%%%%%%%%%%%%%%%%%%%%%%%%%%%%%%%%%%%%%%%%%%%%%%%%%%%%%%%%%%%%%%%%%%%%%%%%%%%%%%%%%%%%%%%%%%%%%%%%%%%%%%%%%%%%%%%%%%%%%%%%%%%%%%%%%%%%%%%%%%%%

%%%%%%%%%%%%%%%%%%%%%%%%%%%%%%%%%%%%%%%%%%%%%%%%%%%%%%%%%%%%%%%%%%%%%%%%%%%%%%%%%%%%%%%%%%%%%%%%%%%%%%%%%%%%%%%%%%%%%%%%%%%%%%%%%%%%%%%%%%%%%%%%%%%%%%%%%%%%%%%%

\begin{figure}
\begin{center}
\includegraphics[scale=0.6]{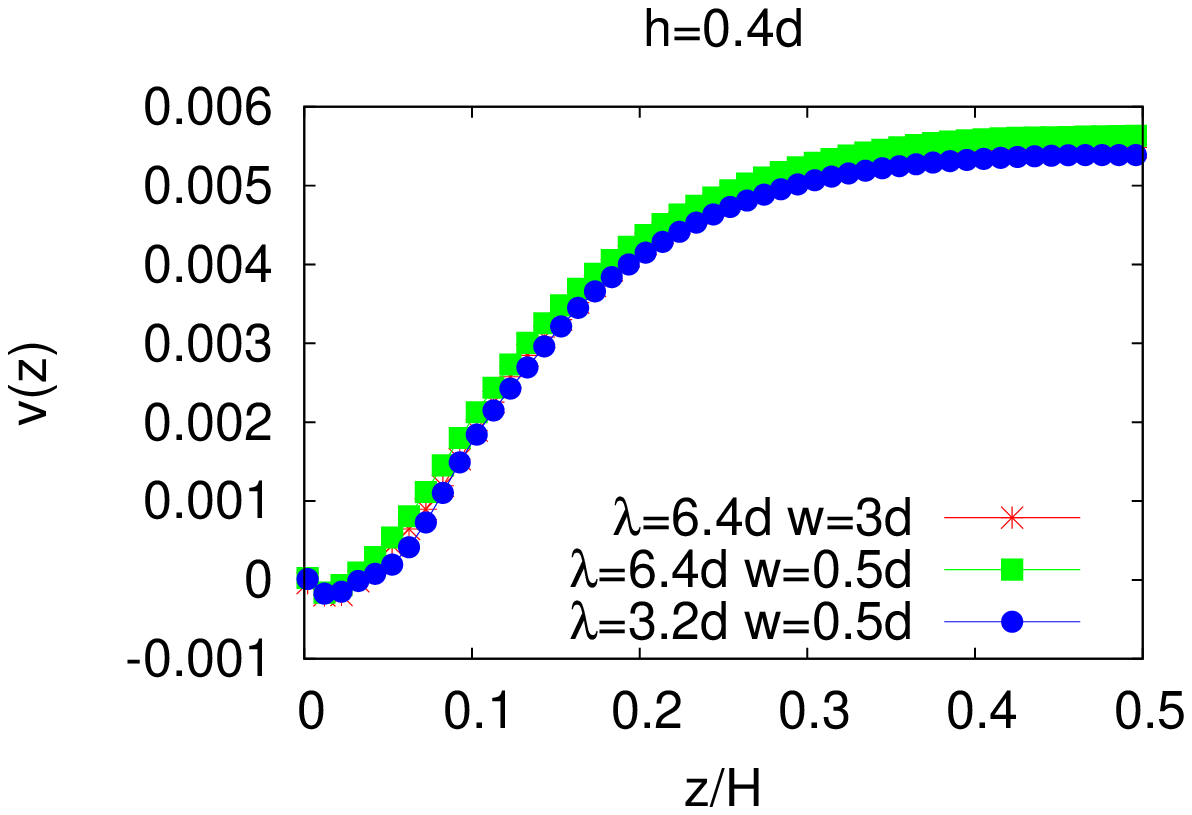}\\
\includegraphics[scale=0.6]{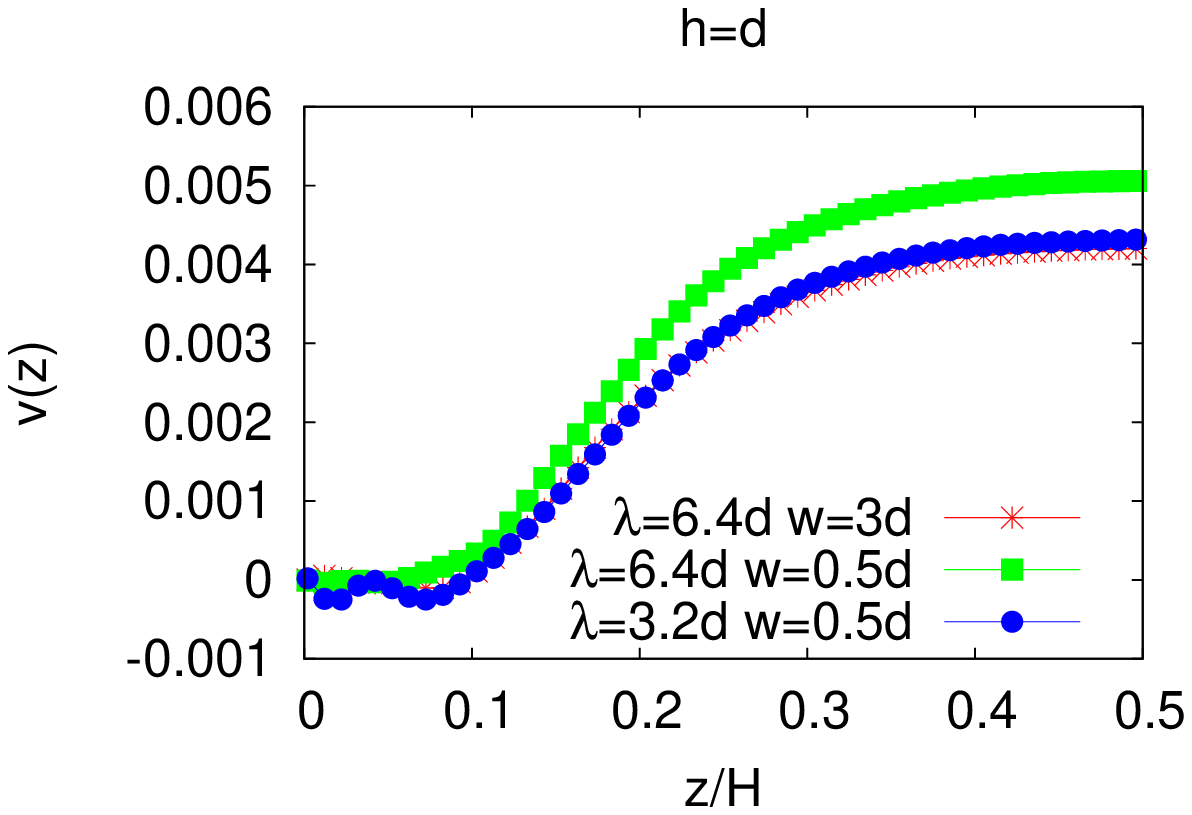}
\caption{Velocity profiles for various post heights $h=0.4\,d$ (top panel) and $h=d$ (bottom panel), for three different couples $(\lambda,w)=(6.4\,d,3\,d)$, $(\lambda,w)=(3.2\,d,0.5\,d)$, $(\lambda,w)=(3.2\,d, 0.5\,d)$. The presence of roughness close to the bottom wall inhibits the macroscopic flow and results in lower mass flow rate through the channel. \label{fig:fixedh}}
\end{center}
\end{figure}

Further insight is gained by the analysis of the various profiles at fixed $h$ and for different couples $(w,\lambda)$, as reported in figure \ref{fig:fixedh}. A comparison of the top ($h=0.4\,d$) and bottom ($h=d$) panels of figure \ref{fig:fixedh} shows that a significant depletion of the mass throughput is achieved when high enough posts are displaced from each other by a short enough distance.

%%%%%%%%%%%%%%%%%%%%%%%%%%%%%%%%%%%%%%%%%%%%%%%%%%%%%%%%%%%%%%%%%%%%%%%%%%%%%%%%%%%%%%%%%%%%%%%%%%%%%%%%%%%%%%%%%%%%%%%%%%%%%%%%%%%%%%%%%%%%%%%%%%%%%%%%%%%%%%%%

%%%%%%%%%%%%%%%%%%%%%%%%%%%%%%%%%%%%%%%%%%%%%%%%%%%%%%%%%%%%%%%%%%%%%%%%%%%%%%%%%%%%%%%%%%%%%%%%%%%%%%%%%%%%%%%%%%%%%%%%%%%%%%%%%%%%%%%%%%%%%%%%%%%%%%%%%%%%%%%%

Taken all together, figures \ref{fig:velocity}-\ref{fig:fixedh}, offer the possibility to put an interesting series of remarks forward. The mass flow rate through the channel depends strongly and in a highly non-trivial way on the geometry and patterning of the wall roughness. The analysis for the velocity profiles reveals that there exists a critical size of the asperities comparable to the droplet diameter which causes inhibition of the flow, and that happens as soon as the inter-post distance $(\lambda-w)$ becomes comparable with the characteristic size of a few droplet diameters, which corresponds to the characteristic size of a T1 plastic rearrangement (typically $\sim 2.5 \div 3 \,d$) \cite{Goyon08,Mansard13,SOFT14} involving the switching of neighboring droplets. It is therefore tempting to establish a link between the results of figures \ref{fig:velocity}-\ref{fig:fixedh} and the mesoscopic plastic dynamics of the material. According to recent theories, experiments and numerical observations \cite{Goyon08,Bocquet09,Geraud13,OURJFM}, such a link can be established with the analysis of the fluidity field $f$ \eqref{eq:proportionality} which is expected to exhibit cooperativity flow behaviour at the macroscopic scales, in agreement with equation \eqref{eq:fluidity}, and with a cooperativity length $\xi$ of the order of a few droplet diameters \cite{Mansard13,SOFT14}.

\subsection{Rate of plastic rearrangements and connection with fluidity field}\label{subsec:rearrange}

\begin{figure}
\begin{center}
\includegraphics[scale=0.6]{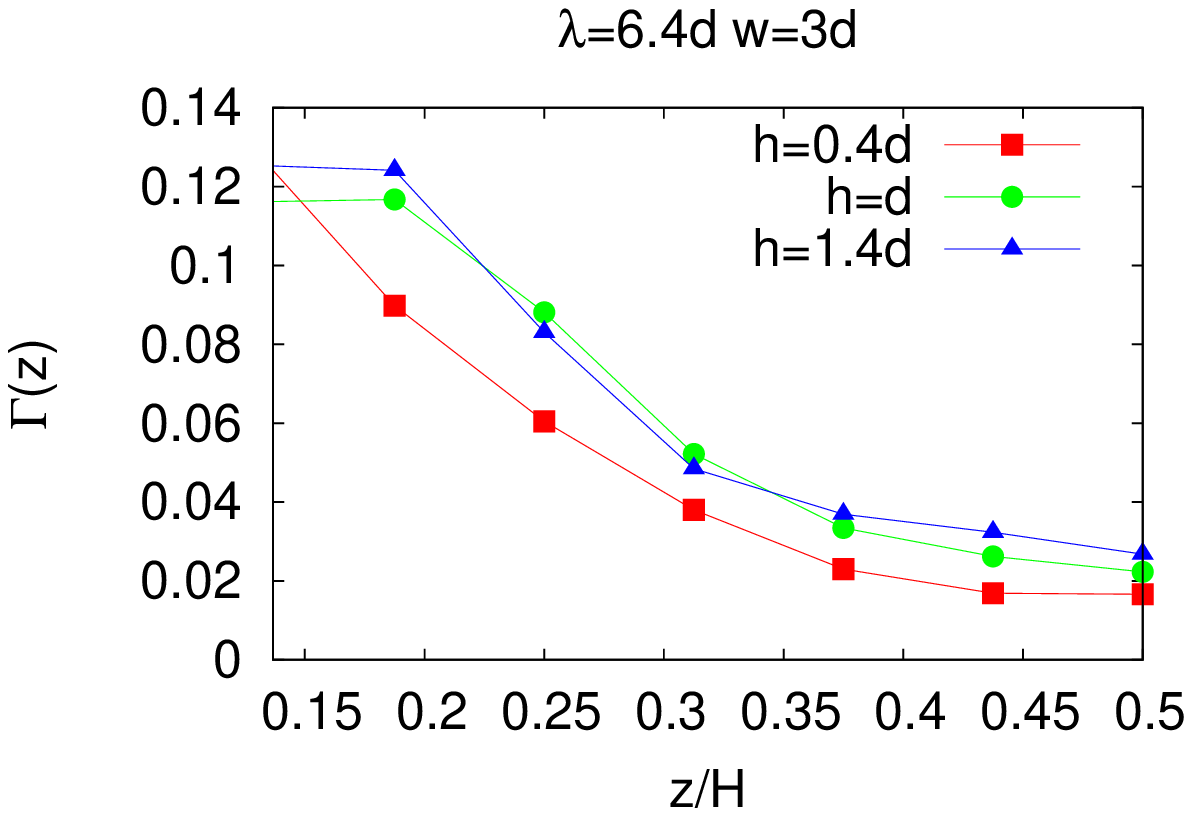}\\
\includegraphics[scale=0.6]{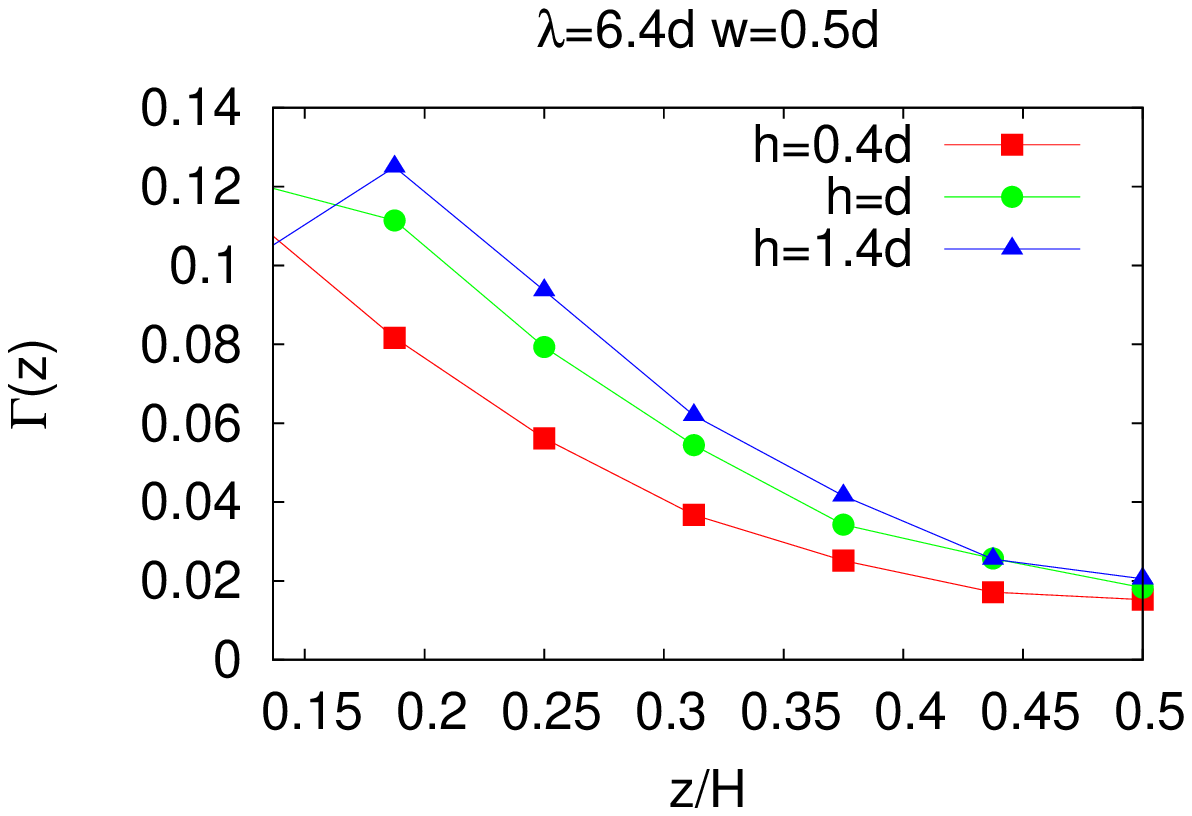}\\
\includegraphics[scale=0.6]{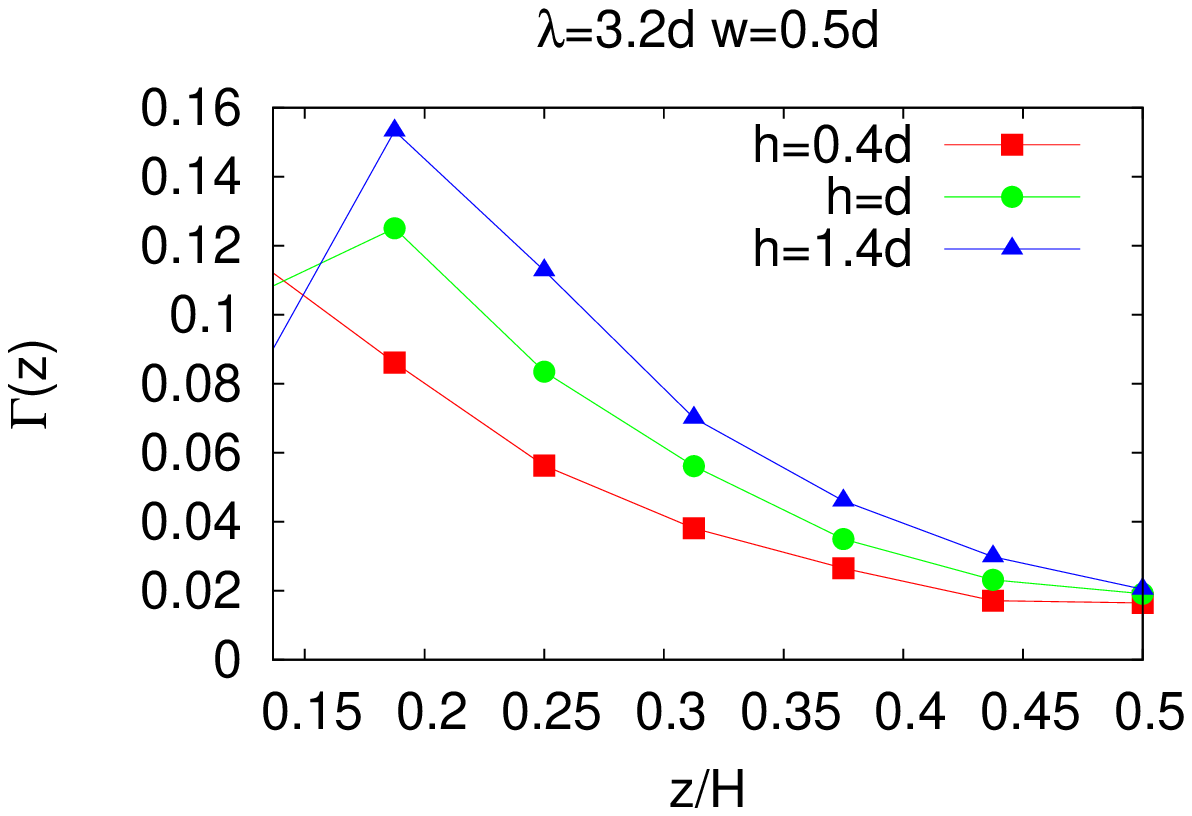}
\caption{Rate of plastic events $\Gamma(z)$ for $(\lambda=6.4\,d, w=3\,d)$ (top panel), $(\lambda=6.4\,d, w=0.5\,d)$ (middle panel) and $(\lambda=3.2\,d, w=0.5\,d)$ (bottom panel). In each panel three sets of data are shown corresponding to different post heights: $h=0.4\,d, d, 1.4\,d$. The $z$ coordinate ranges from $z_0=1.8\,d$ to $H/2$ (mid channel).  \label{fig:T1rate}}
\end{center}
\end{figure}

\begin{figure}[t!]
\begin{center}
\includegraphics[scale=0.3]{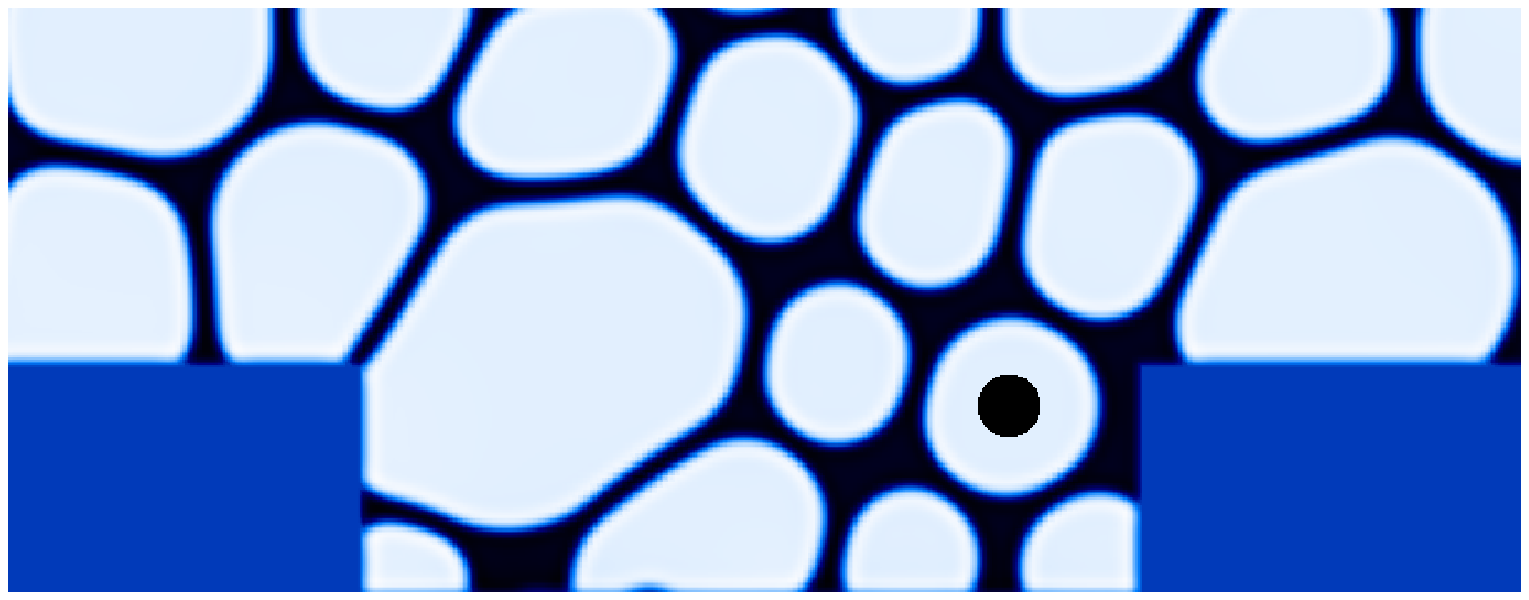}\\ \vspace{.05in}
\includegraphics[scale=0.3]{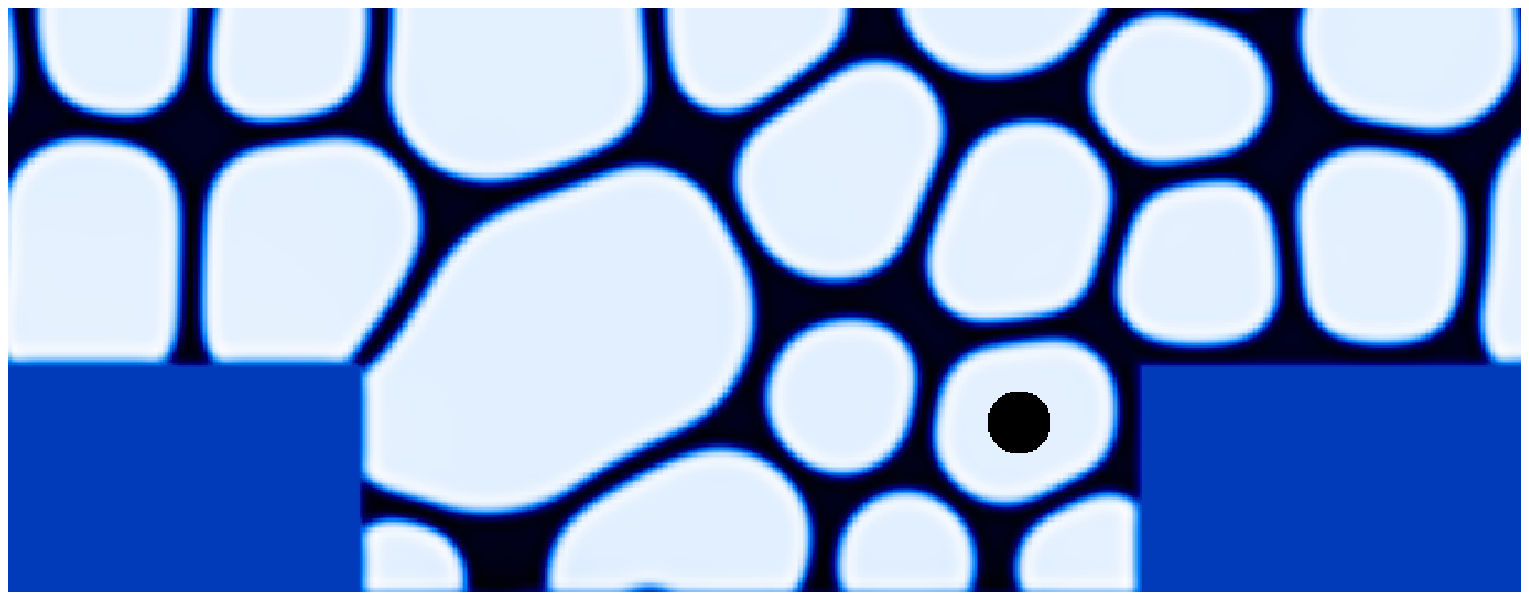}\\ \vspace{.05in}
\includegraphics[scale=0.3]{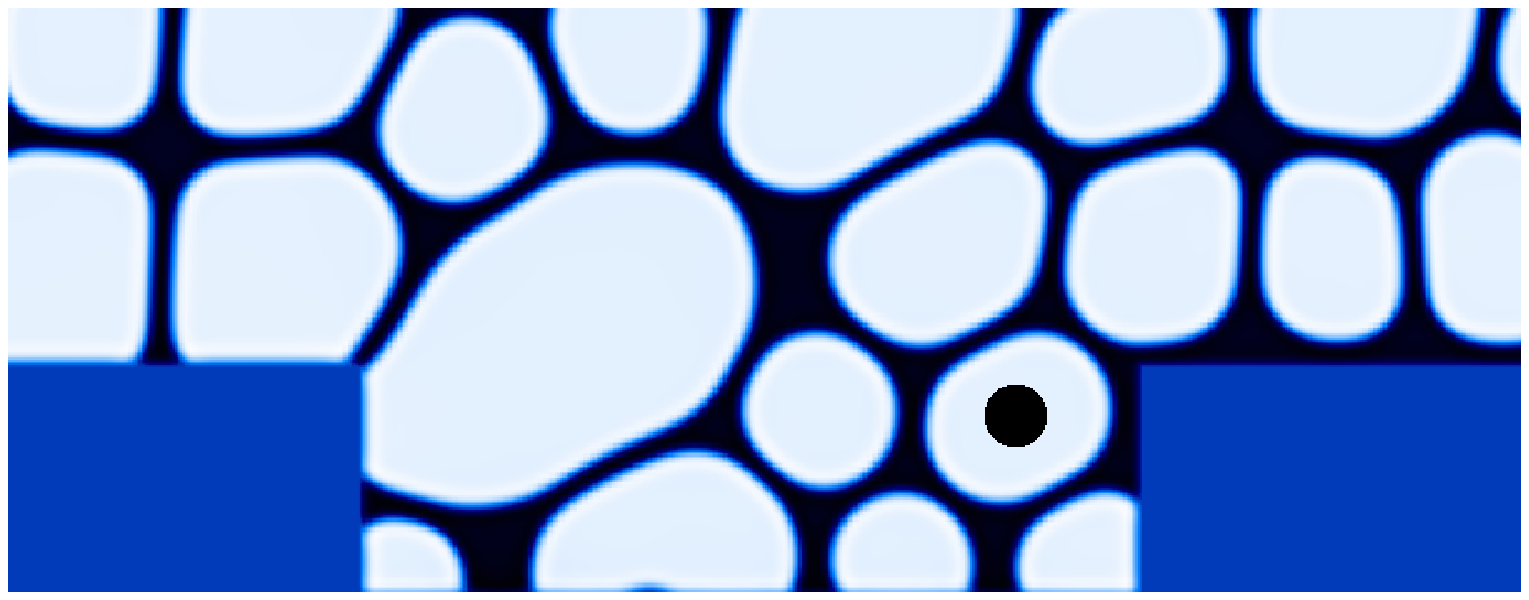}
\caption{Time-ordered (from top to bottom) snapshots of density field of the dispersed phase (light (dark) colors indicate high (low) values of the density) at three instants of time (ordered from top to bottom). Roughness geometrical parameters are: $\lambda=6.4\,d$, $w=3\,d$ and $h=d$. The black bullet indicates the same droplet in the three snapshots. \label{fig:snaps-caged}}
\end{center}
\end{figure}

\begin{figure}[t!]
\begin{center}
\includegraphics[scale=0.3]{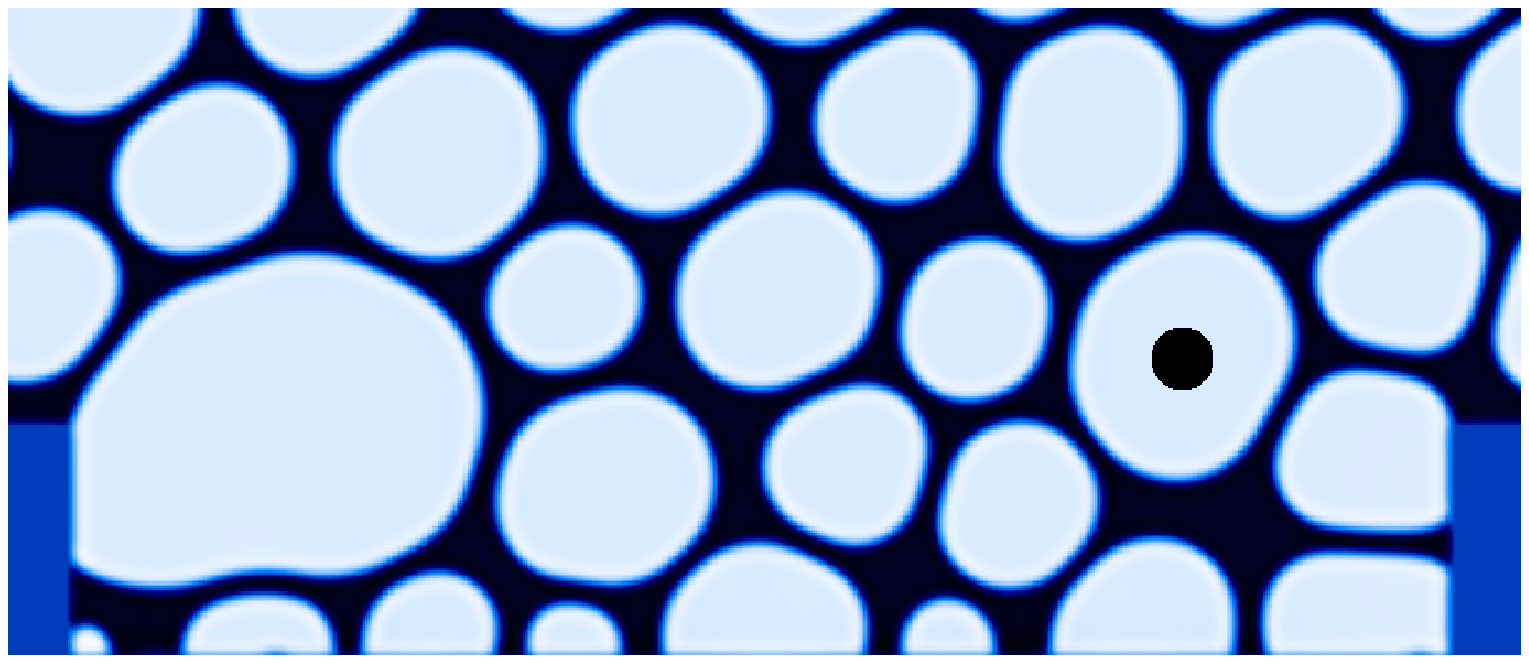}\\ \vspace{.05in}
\includegraphics[scale=0.3]{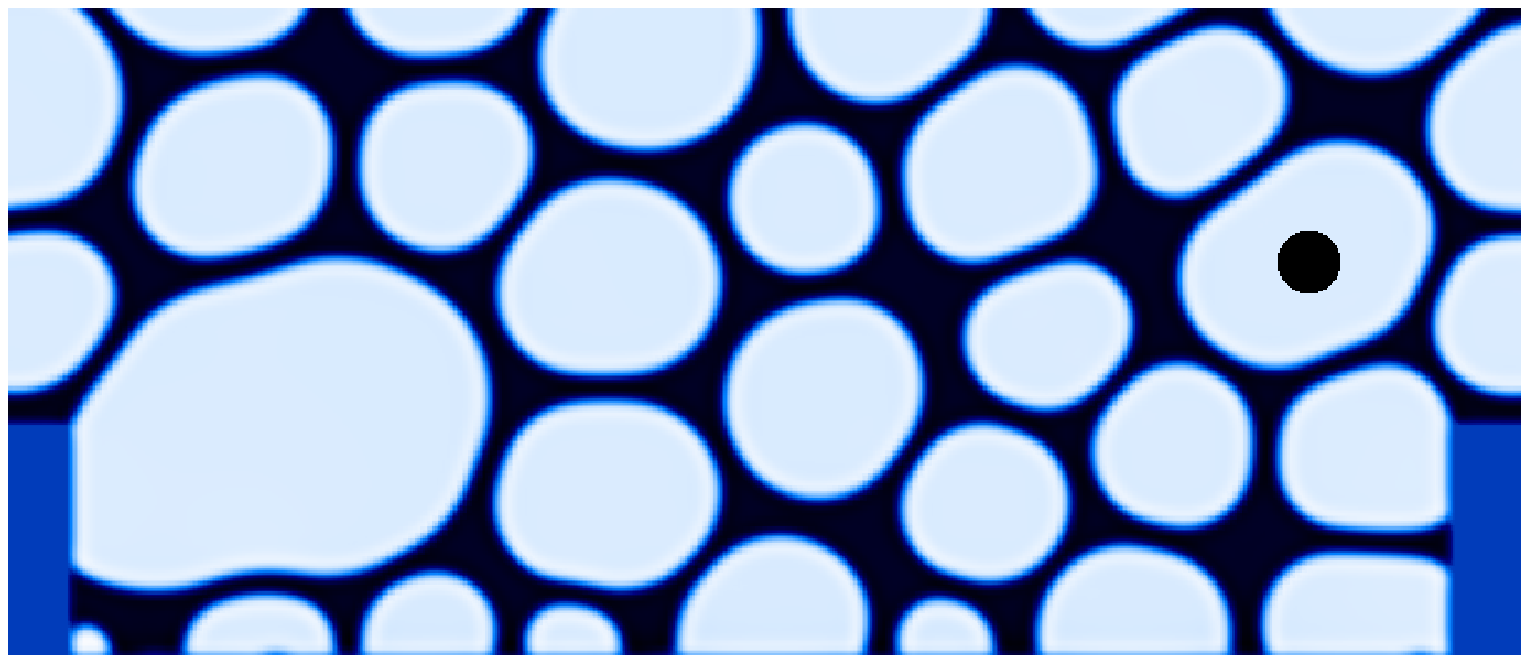}\\ \vspace{.05in}
\includegraphics[scale=0.3]{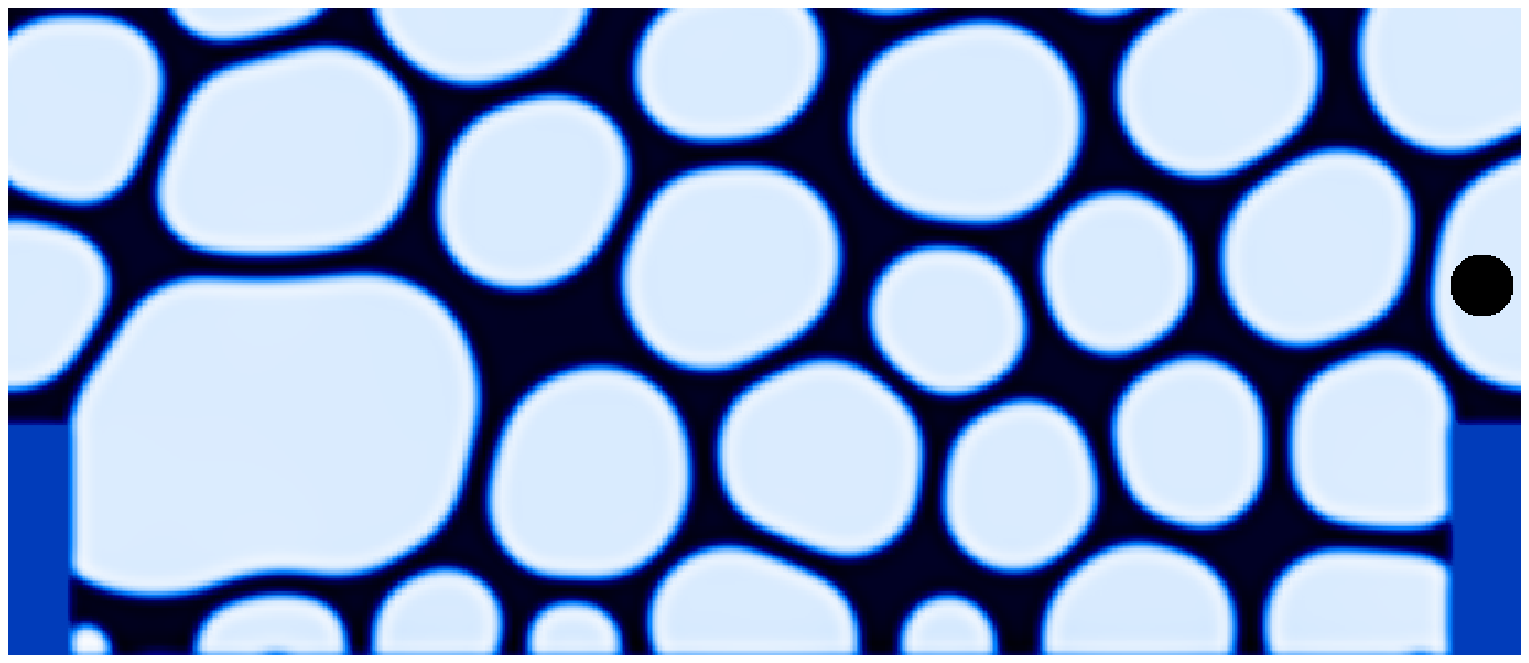}
\caption{Time-ordered (from top to bottom)  snapshots of density field of the dispersed phase (light (dark) colors indicate high (low) values of the density) at three instants of time (ordered from top to bottom). Roughness geometrical parameters are: $\lambda=6.4\,d$, $w=0.5\,d$ and $h=d$. The black bullet indicates the same droplet in the three snapshots. \label{fig:snaps-uncaged}}
\end{center}
\end{figure}

To characterize the various boundary conditions in terms of the rate of plastic events and their connection with the fluidity field, we measured the streamwise-averaged rate $\Gamma (z)$ of T1 rearrangements across the channel. In the spirit of the analysis of the previous section, we compare $\Gamma$'s for various combinations of the wall roughness parameters $\lambda$, $h$ and $w$.  The surface roughness and the distribution of the corrugations have impact on both the ordering of the droplets and the creation of plastic events. In figure \ref{fig:T1rate}, we report the rate of occurrence of such events across the bottom half channel. From top to bottom we have: $(\lambda,w)=(6.4\,d, 3\,d)$, $(\lambda,w)=(6.4\,d, 0.5\,d)$  and $(\lambda,w)=(3.2\,d, 0.5\,d)$; in each panel three sets of data are shown corresponding to different post heights ($h=0.4\,d, d, 1.4\,d$). Overall, roughness is found to ``trigger'' an enhanced rate of plastic rearrangements in the bottom half of the channel. As for the plastic activity between successive posts, it is difficult to infer it: however, for the highest posts ($h=1.4\,d$), we have found a local decrese of the rate of T1 at the edge of the posts, particularly evident when the inter-post distance $(\lambda-w)$ is the smallest (see bottom panel of figure \ref{fig:T1rate}). This suggests that, in order to limit the rotational freedom of groups of neighbouring droplets (i.e. preventing plastic rearrangements), not only the height but also the width of the {\it cages} must fulfill a size criterion. We provide a visual insight of this aspect in figures \ref{fig:snaps-caged}-\ref{fig:snaps-uncaged}, where we show three time-ordered snapshots displaying (figure \ref{fig:snaps-uncaged}) and not displaying (figure \ref{fig:snaps-caged}) caging: by keeping the post height fixed and enlarging the inter-post distance, it is possible to observe that plastic motion is enhanced at larger $(\lambda-w)$. It is worth remarking that, in the experiments of Mansard {\it et al} \cite{Mansard14}, the inter-post distance was set at the value of $3\,d$, roughly the cage size at which
we observe blocking of T1's.

%%%%%%%%%%%%%%%%%%%%%%%%%%%%%%%%%%%%%%%%%%%%%%%%%%%%%%%%%%%%%%%%%%%%%%%%%%%%%%%%%%%%%%%%%%%%%%%%%%%%%%%%%%%%%%%%%%%%%%%%%%%%%%%%%%%%%%%%%%%%%%%%%%%%%%%%%%%%%%%%

%%%%%%%%%%%%%%%%%%%%%%%%%%%%%%%%%%%%%%%%%%%%%%%%%%%%%%%%%%%%%%%%%%%%%%%%%%%%%%%%%%%%%%%%%%%%%%%%%%%%%%%%%%%%%%%%%%%%%%%%%%%%%%%%%%%%%%%%%%%%%%%%%%%%%%%%%%%%%%%%

%%%%%%%%%%%%%%%%%%%%%%%%%%%%%%%%%%%%%%%%%%%%%%%%%%%%%%%%%%%%%%%%%%%%%%%%%%%%%%%%%%%%%%%%%%%%%%%%%%%%%%%%%%%%%%%%%%%%%%%%%%%%%%%%%%%%%%%%%%%%%%%%%%%%%%%%%%%%%%%%

%%%%%%%%%%%%%%%%%%%%%%%%%%%%%%%%%%%%%%%%%%%%%%%%%%%%%%%%%%%%%%%%%%%%%%%%%%%%%%%%%%%%%%%%%%%%%%%%%%%%%%%%%%%%%%%%%%%%%%%%%%%%%%%%%%%%%%%%%%%%%%%%%%%%%%%%%%%%%%%%

%%%%%%%%%%%%%%%%%%%%%%%%%%%%%%%%%%%%%%%%%%%%%%%%%%%%%%%%%%%%%%%%%%%%%%%%%%%%%%%%%%%%%%%%%%%%%%%%%%%%%%%%%%%%%%%%%%%%%%%%%%%%%%%%%%%%%%%%%%%%%%%%%%%%%%%%%%%%%%%%

\begin{figure}[t!]
\begin{center}
\includegraphics[scale=0.6]{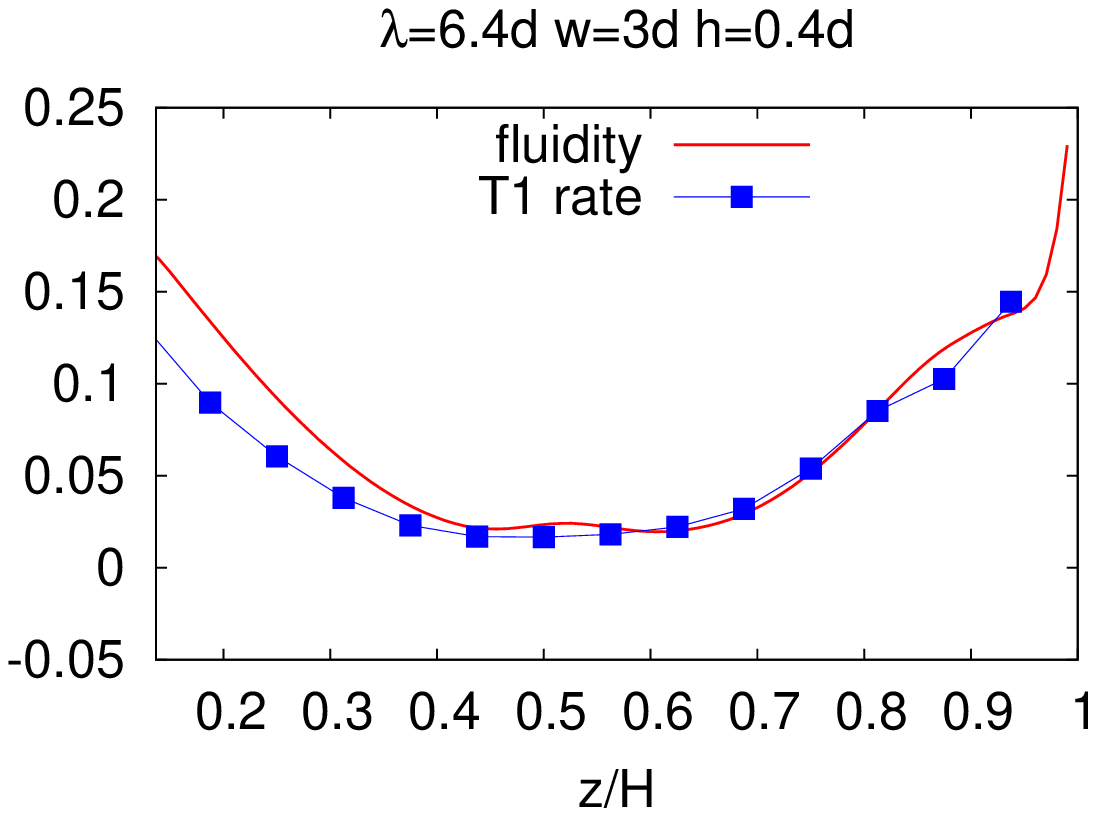}\\
\includegraphics[scale=0.6]{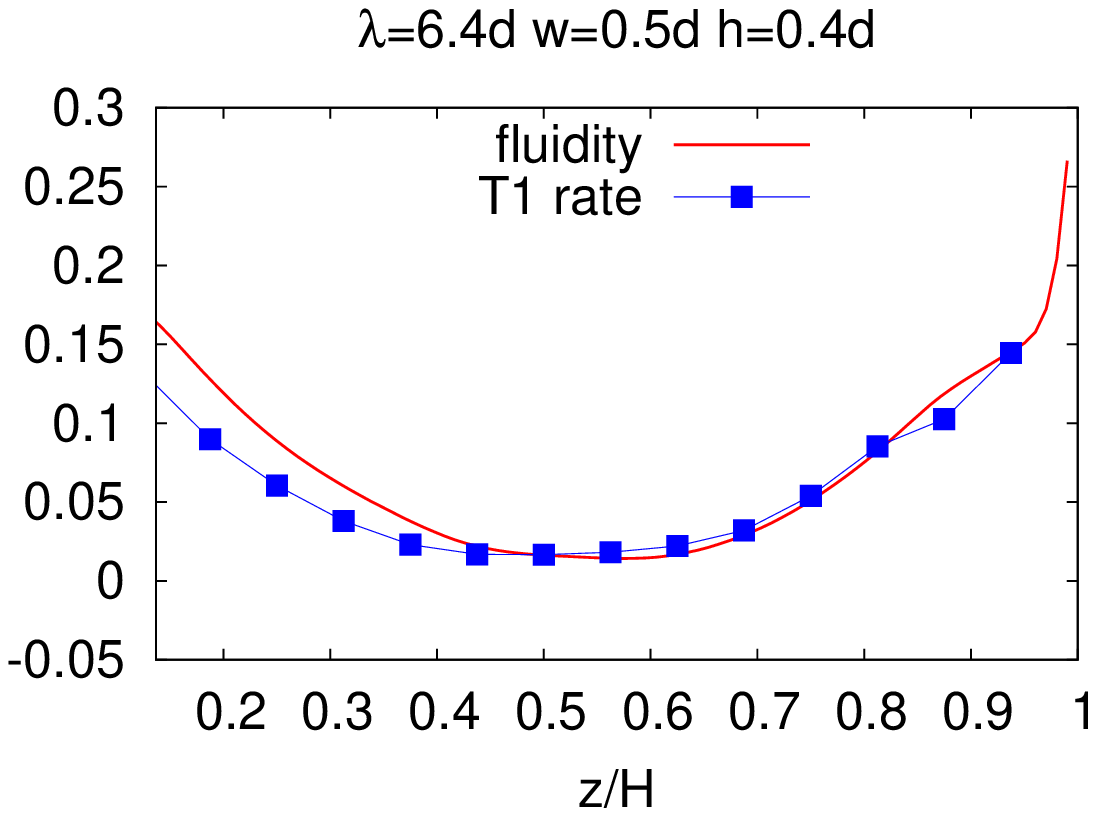}\\
\includegraphics[scale=0.6]{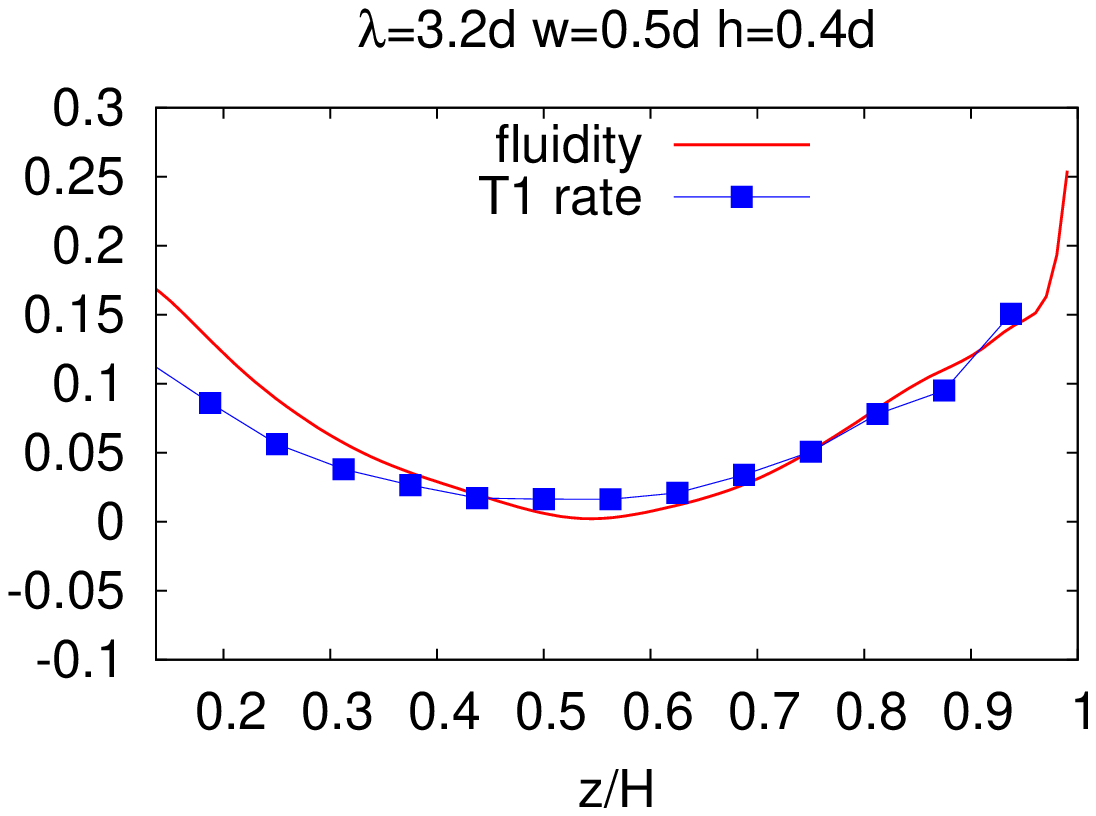}
\caption{Comparison between the fluidity $f=\dot{\gamma}/\sigma$ (red solid line) and rate of T1 events (blue squares) profiles for a fixed post height $h=0.4 \, d$ and three $(\lambda,w)$ couples: $(\lambda,w)=(6.4\,d,3\,d)$ (top panel), $(\lambda,w)=(6.4\,d,0.5\,d)$ (middle panel) and $(\lambda,w)=(3.2\,d,0.5\,d)$ (bottom panel). \label{fig:f+T1-h15}}
\end{center}
\end{figure}

\begin{figure}[t!]
\begin{center}
\includegraphics[scale=0.6]{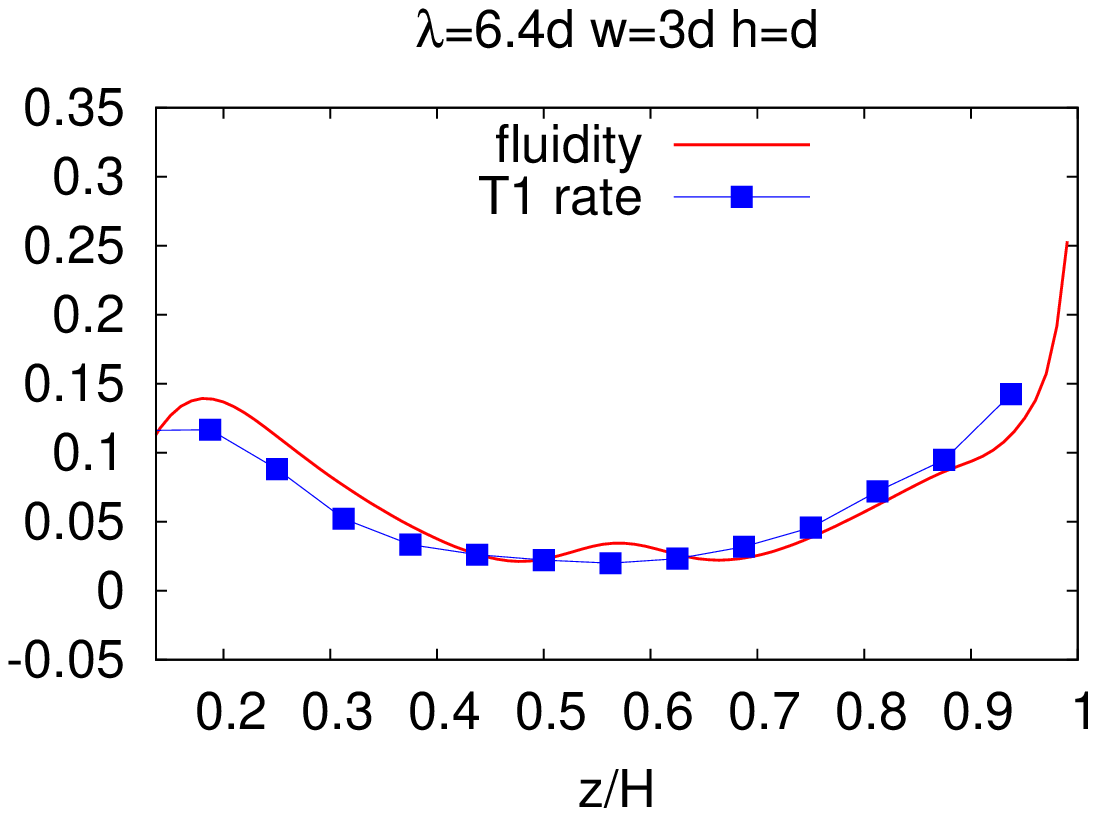}\\
\includegraphics[scale=0.6]{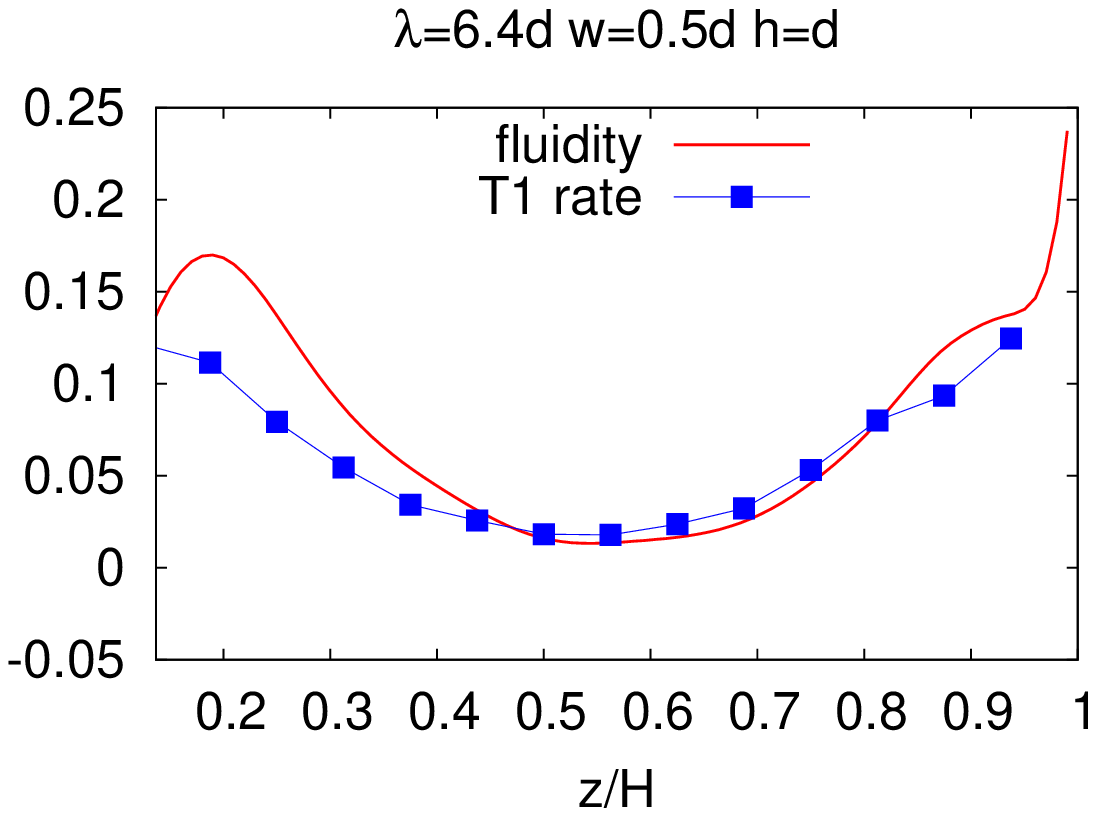}\\
\includegraphics[scale=0.6]{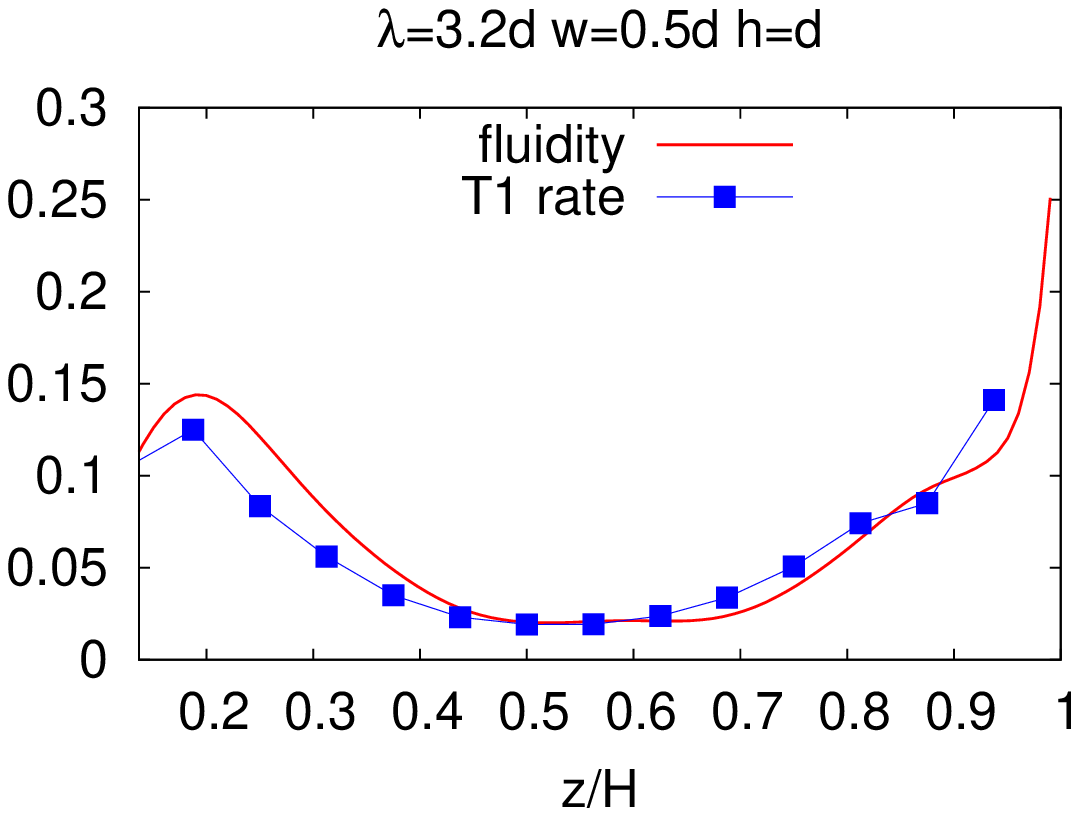}
\caption{Comparison between the fluidity $f=\dot{\gamma}/\sigma$ (red solid line) and rate of T1 events (blue squares) profiles for a fixed post height $h=d$ and three couples $(\lambda,w)$: $(\lambda,w)=(6.4\,d,3\,d)$ (top panel), $(\lambda,w)=(6.4\,d,0.5\,d)$ (middle panel) and $(\lambda,w)=(3.2\,d,0.5\,d)$ (bottom panel). \label{fig:f+T1-h40}}
\end{center}
\end{figure}

Having at hand data on the distribution of T1 rearrangements, we may address the effect of roughness on another crucial issue for the understanding of the microrheology of soft-glassy materials, namely the relation between rate of plastic events and fluidity \cite{Bocquet09}. By combining results from figures \ref{fig:velocity} and \ref{fig:T1rate}, we see that the increase of the plastic activity close to the bottom boundary layer is indeed associated with an enhanced shear rate, therefore supporting the proportionality between fluidity and rate of plastic events (\ref{eq:proportionality}). To be more quantitative, we can also analyze both the fluidity and the rate of plastic events throughout the whole channel. To that purpose, we look at the two quantities for fixed post heights $h=0.4\,d$ (figure \ref{fig:f+T1-h15}) and $h=d$ (figure \ref{fig:f+T1-h40}) and the same three couples of $(\lambda,w)$ considered before. A relevant observation to be made is that in all cases the expected proportionality between $f$ and $\Gamma$ is well satisfied close to the smooth (top) wall; in that region the scaling factor in (\ref{eq:proportionality}) has been fitted to obtain the best matching, and a single value of the constant makes possible a very good description of all profiles.  However, a systematic mismatch emerges close to the bottom wall, with the fluidity that overestimates the rate of plastic events. One could argue that the bottom boundary layer is characterized by a local elastic modulus $G_0$ smaller than that close to the top wall, thus explaining the mismatch. A visual inspection of snapshots from simulations in the steady state (see figure \ref{fig:sketch}) shows, indeed, that the grooves trigger the formation of droplets with a characteristic size slightly larger than those near the top wall; consequently, bottom droplets are ``less elastic'' (more easily deformable) than those on top. Another possible source for the mismatch may arise from the fact that the bottom boundary layer is characterized by a modulated plastic activity: regions with posts (i.e. zero plastic activity) alternate with ``holes'' where droplets are present and may rearrange. This mixed boundary condition \cite{SbragagliaProsperetti} for the plastic activity breaks the streamwise (statistical) invariance of the problem \cite{Einzel,Panzer} and suggests the need of a modification of relation (\ref{eq:proportionality}) by an {\it effective} proportionality factor (smaller than one) dependent on the geometrical details, that may improve the agreement between the measured fluidity and the rate of plastic events. Remarkably, the agreement between the fluidity and the rate of plastic events is, instead, good enough for the two cases with ``cage effect''(top and bottom panels of figure \ref{fig:f+T1-h15}): this may be taken as a further indication that, for a suitable choice of the inter-post distance, droplets switching close to the bottom boundary layer is so inhibited that the mixed character of the boundary condition for the plastic activity is lost as well. We hasten to add, however, that these interpretations, although pretty plausible, surely need consolidation through further systematic investigations, both numerically and, possibly, analytically.

%%%%%%%%%%%%%%%%%%%%%%%%%%%%%%%%%%%%%%%%%%%%%%%%%%%%%%%%%%%%%%%%%%%%%%%%%%%%%%%%%%%%%%%%%%%%%%%%%%%%%%%%%%%%%%%%%%%%%%%%%%%%%%%%%%%%%%%%%%%%%%%%%%%%%%%%%%%%%%%%

%%%%%%%%%%%%%%%%%%%%%%%%%%%%%%%%%%%%%%%%%%%%%%%%%%%%%%%%%%%%%%%%%%%%%%%%%%%%%%%%%%%%%%%%%%%%%%%%%%%%%%%%%%%%%%%%%%%%%%%%%%%%%%%%%%%%%%%%%%%%%%%%%%%%%%%%%%%%%%%%

%%%%%%%%%%%%%%%%%%%%%%%%%%%%%%%%%%%%%%%%%%%%%%%%%%%%%%%%%%%%%%%%%%%%%%%%%%%%%%%%%%%%%%%%%%%%%%%%%%%%%%%%%%%%%%%%%%%%%%%%%%%%%%%%%%%%%%%%%%%%%%%%%%%%%%%%%%%%%%%%

\section{Conclusions and perspectives}\label{sec:conclusions}

Based on lattice Boltzmann simulations of emulsion droplets, we have characterized the impact of geometrical roughness on Poiseuille flow profiles of a dense emulsion above the jamming point. Specifically, we have performed numerical simulations of a collection of closely packed droplets in a channel where one wall is decorated with a periodic array of posts with variable wavelength, width and height. The independent analysis of plastic rearrangements in the flowing material allows to establish a link between the velocity profiles and the mesoscopic plastic dynamics. Actually, plastic rearrangements and the emergence of their spatial correlations induce a cooperativity flow behavior whose effect is pronounced in confined situations like those analyzed in this paper. More in detail, based on the work of Goyon {\it et al.} \cite{Goyon08,Goyon10,Jop12} we have explored the relation between the fluidity, defined as the ratio between shear rate and shear stress, and the rate of plastic events. Our results confirm the expected proportionality between the two quantities throughout the channel, although surface roughness is found to slightly spoil the expected behaviour, especially close to the rough boundary layer where modulated boundary conditions emerge for the plastic activity. \\
In the general perspective of exploring the role of heterogeneous boundary conditions, one may also wonder about the effects induced by the chemical patterning. To this aim, preliminary numerical simulations reveal that such an effect is less pronounced than that induced by roughness, probably because the droplets deformation is smaller close to the boundary layer. Upon increasing the local capillary number, however, droplets may be more easily detached from the wall, therefore preventing the formation of a boundary layer where jammed droplets can move upon the effect of the stick-slip dynamics \cite{prl13} associated with the motion of the contact lines. This is surely an interesting point worth of future investigations, both numeric and analytic. We argue that our work may enhance the interest in discussing the emergence of correlations in plastic rearrangements close to boundary conditions which are peculiar of many applied microfluidics geometries \cite{Seeman}.

\section{Acknowledgements}
The authors kindly acknowledge funding from the European Research Council under the European Community's Seventh Framework Programme (FP7/2007-2013)/ERC Grant Agreement No. 279004.

\end{document}